\documentclass{elsart3p}

\usepackage{graphics}
\usepackage{graphicx}
\usepackage{amssymb}

\newcommand{\D}{\delta }

\def\vq{{\bf q}}

\newcommand{\fig}[1]{Fig.~\ref{#1}}
\newcommand{\be}{\begin{equation}}
\newcommand{\ee}{\end{equation}}

\begin{document}

\begin{frontmatter}

%titles, authors and addresses


\title{Theory of the in-plane anisotropy of magnetic excitations in 
YBa$_{2}$Cu$_{3}$O$_{6+y}$}


\author{Hiroyuki Yamase and Walter Metzner}

\address{Max-Planck-Institute for Solid State Research, 
Heisenbergstrasse 1, D-70569 Stuttgart, Germany}

\begin{abstract}
A pronounced $xy$-anisotropy was observed in recent neutron scattering
experiments for magnetic excitations in untwinned 
YBa$_{2}$Cu$_{3}$O$_{6+y}$ (YBCO). 
The small anisotropy of the bare band structure due to
the orthorhombic crystal symmetry seems to be enhanced by correlation
effects. A natural possibility is that the system is close to a
Pomeranchuk instability associated with a $d$-wave Fermi surface 
deformation ($d$FSD).
We investigate this possibility in the bilayer $t$-$J$ model within
a self-consistent slave-boson mean-field theory.
We show that the $d$FSD correlations drive a pronounced $xy$-anisotropy
of magnetic excitations at low doping and at relatively high
temperatures, providing a scenario for the observed $xy$-anisotropy 
in optimally doped as well as underdoped YBCO, including the pseudogap 
phase. 
\end{abstract}

\end{frontmatter}

% main text
%\section{Title of the first section}
%\label{labelOfFirstSection}
The interpretation of neutron scattering data for high-$T_{c}$ 
cuprates within a spin-charge stripe scenario has attracted
much interest\cite{tranquada95}. 
It predicts a satellite signal at either $\vq=(\pi,\pi\pm 2\pi\eta)$ or 
$(\pi\pm 2\pi\eta,\pi)$, depending on the direction of the stripes. 
Recently Hinkov {\it et al.}\cite{hinkov0406} performed 
neutron scattering experiments for untwinned YBa$_{2}$Cu$_{3}$O$_{6+y}$ 
to check this scenario. 
The untwinned crystals  
are purely orthorhombic and thus it is expected that 
the stripe direction is determined uniquely by the orthorhombicity, 
leading to only one satellite signal. 
However, they observed two satellite 
signals at both $\vq=(\pi,\pi\pm 2\pi\eta_{y})$ and 
$(\pi\pm 2\pi\eta_{x},\pi)$, with $\eta_{x}\ne \eta_{y}$ and 
different intensity, 
and concluded that the stripe scenario did not work. 
Then how can we understand this $xy$-anisotropy of 
magnetic excitations? 
Our scenario is that it comes from 
electron correlation effects associated with a 
$d$-wave Fermi surface deformation ($d$FSD). 

The $d$FSD was first discussed in the $t$-$J$\cite{yamase00} 
and the Hubbard\cite{halboth00} model: 
the Fermi surface expands 
along the $k_{x}$ direction and shrinks along the $k_{y}$ direction, 
or vice versa.  
The $d$FSD is driven by forward scattering interactions of 
quasi-particles (Pomeranchuk instability).  
The $d$FSD competes with the $d$-wave superconductivity ($d$SC). 
In the slave-boson mean-field analysis of the $t$-$J$ 
model\cite{yamase00}, the dominant instability is the $d$SC and the 
spontaneous $d$FSD does not occur. 
However, the system still has appreciable $d$FSD correlations, 
which yield an enhanced anisotropy of a 
renormalized band structure in the presence of an external 
anisotropy such as orthorhombicity of a lattice\cite{yamase00}. 

We explore this $d$FSD correlation effect and 
compute magnetic excitations in the bilayer 
$t$-$J$ model in the slave-boson mean-field scheme with a small 
$xy$-anisotropy in $t$ and $J$. 
The magnetic excitations contain the even and the odd channel. 
Comprehensive results for the odd channel as well as our detailed 
formalism are presented in 
Ref.\cite{yamase06}. We focus on the even channel in this paper 
and take the same model parameters as in Ref.\cite{yamase06}. 

\begin{figure}[t]
%\begin{figure}[!ht]
\begin{center}
\includegraphics[width=0.37\textwidth]{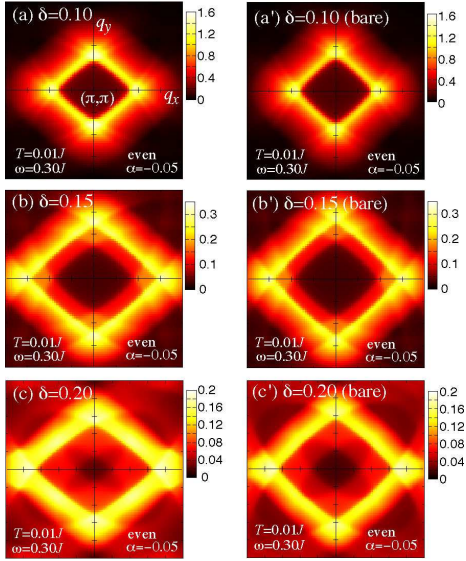}
\end{center}
\caption{(Color online) 
$\vq$ maps of magnetic excitation spectra for a sequence of 
$\D$ for $T=0.01J$ and $\omega=0.30J$ 
in the even channel; $5\% (10\%)$ anisotropy is introduced in $t(J)$; 
the right-hand panels show a bare anisotropy effect, namely 
without $d$FSD correlations; 
$\vq$ is  scanned in $0.6\pi \leq q_{x}, q_{y} \leq 1.4\pi$ 
except for the panels for $\D=0.20$ where 
$0.5\pi\leq q_{x}, q_{y} \leq 1.5\pi$.}
\label{qmapde}
\end{figure}

$\vq$ maps of magnetic excitation spectra  
are shown in \fig{qmapde} 
for a sequence of hole-doping rates $\delta$ at low $T$. 
The left-hand panels are self-consistent results while 
the right-hand panels are corresponding non-self-consistent results 
where $d$FSD correlations are switched off, keeping just the
bare input anisotropy. 
The strong spectral weight forms a diamond shaped distribution around 
$\vq=(\pi,\pi)$ with prominent weight at 
$\vq=(\pi,\pi\pm 2\pi\eta_{y})$ and $(\pi\pm 2\pi\eta_{x},\pi)$. 
A robust measure of the anisotropy of magnetic excitations 
is given by $\Delta\eta=\eta_{x}-\eta_{y}$\cite{yamase06}.  
Although the input anisotropy is fixed, we can read off 
appreciable $\delta$ dependence of $\Delta\eta$ from \fig{qmapde}. 
In particular, the bare anisotropy effect (right-hand panels) 
becomes less effective for lower $\D$, where 
an almost isotropic distribution is seen. 
The $d$FSD correlations, on the other hand, enhance the bare anisotropy 
effect and yield a sizable anisotropy even for low $\D$. 

\begin{figure}[t]
%\begin{figure}[!ht]
\begin{center}
\includegraphics[width=0.37\textwidth]{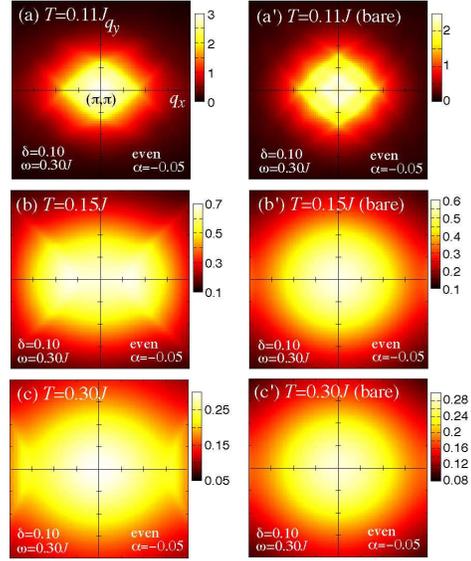}
\end{center}
\caption{(Color online) 
$\vq$ maps of magnetic excitation spectra for a sequence of $T$ 
in $0.6\pi\leq q_{x}, q_{y} \leq 1.4\pi$ for   
$\D=0.10$ and $\omega=0.30J$ in the even channel; 
$5\% (10\%)$ anisotropy is introduced in $t(J)$; 
right-hand panels show a bare anisotropy effect; 
the $d$-wave singlet is realized below 
$T_{\rm RVB}=0.133J (0.139J)$ 
%$T<0.133J (0.139J)$ 
in the left-hand (right-hand) cases.} 
\label{qmapTe}
\end{figure}

The importance of the $d$FSD correlations appears also 
in the $T$ dependence. 
The diamond shaped distribution at low $T$ (\fig{qmapde}) is rather 
robust against $T$ although 
the spectral weight around $\vq=(\pi,\pi)$ increases with $T$. 
Such spectral weight finally 
becomes dominant [\fig{qmapTe}(a)] and a 
pronounced $xy$-anisotropy appears. 
Figure~\ref{qmapTe} shows a comparison between results including 
$d$FSD correlations (left-hand panels) and results based on the 
bare anisotropy only (right-hand panels), for a sequence of $T$. 
We see that $d$FSD correlations drive a pronounced anisotropy 
for temperatures around $T_{\rm RVB}$ (corresponding to the
pseudogap temperature) while they become less effective for 
higher $T$. 

We have seen that the $d$FSD correlations lead to a pronounced 
anisotropy of magnetic excitations at low $\D$ and at relatively 
high $T$, which are common features in both the odd\cite{yamase06} 
and the even channel. 
Anisotropic magnetic excitations are observed 
in YBa$_{2}$Cu$_{3}$O$_{6.6}$\cite{hinkov0406} 
and YBa$_{2}$Cu$_{3}$O$_{6.5}$\cite{stock04},  
and in the pseudogap phase\cite{hinkov0406}. 
These data can not be understood by the bare anisotropy effect only 
(see right-hand panels in Figs.~\ref{qmapde} and \ref{qmapTe}), 
implying the importance of $d$FSD correlations.

\end{document}